\documentclass[12pt]{article}
\usepackage{amsmath}
\usepackage{amssymb}
\usepackage{graphicx}
\usepackage{enumerate}
\usepackage{natbib}
\usepackage{color}
\usepackage{float}
\usepackage{url} 
\usepackage{xr}

\bibpunct{(}{)}{,}{autheryear}{,}{;}

\newcommand{\blind}{1}

\newcommand*{\indep}{%
	\mathbin{%
		\mathpalette{\@indep}{}%
	}%
}

\newtheorem{assumption}{Assumption}
\newtheorem{theorem}{Theorem}

\newtheorem{remark}{Remark}
\newtheorem{example}{Example}

\newcommand{\pr}{P}

\newcommand{\R}{\mathbb{R}}
\newcommand{\de}{\mathrm{d}}
\newcommand{\T}{\mathrm{\scriptscriptstyle T}}
\newcommand{\logit}{\text{logit}}

\newcommand{\ipw}{\mathrm{IPW}}

\newcommand{\reg}{\mathrm{reg}}

\newcommand{\ora}{\mathrm{ora}}

\newcommand{\Cal}{\mathrm{cal}}
\newcommand{\sign}{\mathrm{sign}}
\newcommand{\dr}{\mathrm{dr}}
\newcommand{\A}{\mathcal{A}}
\newcommand{\B}{\mathcal{B}}
\newcommand{\C}{\mathcal{C}}
\newcommand{\J}{\mathcal{J}}
\newcommand{\N}{\mathcal{N}}
\newcommand{\U}{\mathcal{U}}
\newcommand{\F}{\mathcal{F}}

\newcommand{\p}{\mathrm{p}}
\newcommand{\M}{\mathcal{M}}
\newcommand{\pipw}{\text{p-ipw}}
\newcommand{\preg}{\text{p-reg}}
\newcommand{\pee}{\text{p-dr}}
\newcommand{\naive}{\mathrm{naive}}

\begin{document}

\def\spacingset#1{\renewcommand{\baselinestretch}%
{#1}\small\normalsize} \spacingset{1}


\if1\blind
{
  \title{\bf Doubly Robust Inference when Combining Probability and Non-probability Samples with High-dimensional Data}
  \author{Shu Yang$^1$ \thanks{{syang24@ncsu.edu}, Department of Statistics, North Carolina 27695, U.S.A.},
  	 Jae Kwang Kim$^2$, and Rui Song$^1$\\
    $^{1}$ Department of Statistics, North Carolina State University,\\
    $^2$Department of Statistics, Iowa State University}
\date{}
  \maketitle
} \fi

\if0\blind
{
  \bigskip
  \bigskip
  \bigskip
  \begin{center}
    {\LARGE\bf Doubly Robust Inference when Combining Probability and Non-probability Samples with High-dimensional Data}
\end{center}
  \medskip
} \fi

\bigskip

\noindent%
\textbf{Summary.} We consider integrating a non-probability sample with a probability sample which provides high-dimensional representative covariate information of the target population. We propose a two-step approach for variable selection and finite population inference. In the first step, we use penalized estimating equations with folded-concave penalties to select important variables and show the selection consistency for general samples. In the second step, we focus on a doubly robust estimator of the finite population mean and re-estimate the nuisance model parameters by minimizing the asymptotic squared bias of the doubly robust estimator. This estimating strategy mitigates the possible first-step selection error and renders the doubly robust estimator root-$n$ consistent if either the sampling probability or the outcome model is correctly specified.


\noindent%
\vfill

\newpage
\spacingset{1} 

\section{Introduction}

Probability sampling is regarded as the gold-standard in survey statistics
for finite population inference. Fundamentally, probability samples
are selected under known sampling designs and therefore are representative
of the target population. However, many practical challenges
arise in collecting and analyzing probability sample data such as
cost, time duration, and increasing non-response rates \citep{keiding2016perils}.
As the advancement of technology, non-probability samples become
increasingly available for research purposes, such as remote sensing
data, web-based volunteer samples, etc. Although non-probability samples do not
contain information on the sampling mechanism, they provide rich information
about the target population and can be potentially helpful for finite
population inference. These complementary features of probability
samples and non-probability samples raise the question of whether
it is possible to develop data integration methods that leverage the
advantages of both data sources.

Existing methods for data integration can be categorized into three
types. The first type is the so-called propensity score adjustment
\citep{rosenbaum1983central}. In this approach, the probability of
a unit being selected into the non-probability sample, which is referred
to as the propensity or sampling score, is modeled and estimated for
all units in the non-probability sample. The subsequent adjustments,
such as propensity score weighting or stratification, can then be
used to adjust for selection biases; see, e.g., \citet{lee2009estimation,valliant2011estimating,elliott2017inference}
and \citet{chen2018doubly}. \citet{stuart2011use,stuart2015assessing}
and \citet{buchanan2018generalizing} use propensity score weighting
to generalize results from randomized trials to a target population.
\citet{o2014generalizing} propose propensity score stratification
for analyzing a non-randomized social experiment. One notable disadvantage
of the propensity score methods is that they rely on an explicit propensity
score model and are biased and highly variable if the model is misspecified
\citep{kang2007demystifying}. The second type uses calibration weighting
\citep{deville1992calibration,kott2006using,chen2018model,chen2019calibrating}. This technique calibrates
auxiliary information in the non-probability sample with that in the
probability sample, so that after calibration the weighted distribution
of the non-probability sample is similar to that of the target population
\citep{disogra2011calibrating}. The third type is mass imputation,
which imputes the missing values for all units in the probability
sample. In the usual imputation for missing data analysis, the respondents
in the sample constitute a training dataset for developing an imputation
model. In the mass imputation, an independent non-probability sample
is used as a training dataset, and imputation is applied to all units
in the probability sample; see, e.g., \citet{breidt1996two}, \citet{rivers2007sampling},
\citet{kim2012combining}, \citet{chipperfield2012combining}, \citet{bethlehem2016solving},
and \citet{yang2018integration}.

Let $X\in\R^{p}$ be a vector of auxiliary variables (including an
intercept) that are available from two data sources, and let $Y$
be a general-type study variable of interest. We consider combining a probability
sample with $X$, referred to as Sample A, and a non-probability sample
with $(X,Y)$, referred to as Sample B, to estimate $\mu$ the population
mean of $Y$. Because the sampling mechanism of a non-probability
sample is unknown, the target population quantity is not identifiable
in general. Researchers rely on an identification strategy that requires
a non-informative sampling assumption imposed on the non-probability
sample. To ensure this assumption holds, researchers should control
for all covariates that are predictors of both sampling and the outcome
variable. In practice, subject matter experts will recommend a rich
set of potential useful variables but will not identify the exact
variables to adjust for. In the presence of many auxiliary variables,
variable selection becomes important, because existing methods may
become unstable or even infeasible, and irrelevant auxiliary variables
can introduce a large variability in estimation. There is a large
literature on variable selection methods for prediction, but little
work on variable selection for data integration that \textcolor{black}{can
	successfully recognize the strengths and the limitations of each data
	source and utilize all information captured for finite population
	inference.} \citet{gao2017data} propose a pseudo-likelihood approach
to combining multiple non-survey data with high dimensionality; this
approach requires all likelihoods be correctly specified and therefore
is sensitive to model misspecification.
\citet{chen2018model} propose a model-assisted calibration approach using LASSO; this approach relies on a correctly specified outcome model.
Up to our knowledge, robust inference has not been addressed in the context
of data integration with high-dimensional data.

We propose a doubly robust variable selection and estimation strategy
that harnesses the representativeness of the probability sample and
the outcome  information in the non-probability sample.
The double robustness entails that the final estimator is consistent
for the true value if either the probability of selection into the non-probability
sample, referred to as the sampling score, or the outcome model is
correctly specified, not necessarily both (a double robustness condition); see, e.g.,
\citet{bang2005doubly,tsiatis2007semiparametric,cao2009improving}, and \citet{han2013estimation}.
To handle potentially high-dimensional covariates, our strategy separates
the variable selection step and the estimation step for finite
population mean to achieve two different goals.

In the first step, we select a set of variables that are important
predictors of either the sampling score or the outcome model by penalized
estimating equations. Following most of the empirical literature,
we assume the sampling score follows a logistic regression model with
the unknown parameter $\alpha\in\R^{p}$ and the outcome follows a
generalized linear model (accommodating different types of the outcome)
with the unknown parameter $\beta\in\R^{p}$.
Importantly, we separate  the estimating equations for $\alpha$ and
$\beta$ in order to achieve stability in variable selection
under the double robustness condition.
Specifically, we construct the estimating
equation for $\alpha$ by calibrating the weighted average of $X$ from Sample
B, weighted by the inverse of the sampling score, to the design weighted average of $X$ from Sample A (i.e., a design
estimate of population mean of $X$).
We construct the estimating
equation for $\beta$ by
minimizing the standard least squared error loss under the outcome model.
To  establish
the selection properties,
we consider the ``large $n$, diverging $p$'' framework.
To the best of our knowledge,
the asymptotic properties of penalized estimating estimation based
on survey data have not been studied in the literature. Our major
technical challenge is that under the finite population framework,
the sampling indicator of Sample A may not be independent	even under simple random sampling. To overcome this challenge, we
construct martingale random variables with a weak dependence that
allows applying Bernstein inequality. This construction is innovative
and crucial in establishing our new selection consistency result.

 In the second step, we consider a doubly robust estimator of $\mu$, $\widehat{\mu}_{\dr}(\widehat{\alpha},\widehat{\beta})$, and  re-estimate
 $({\alpha},{\beta})$
 based on the joint set of covariates selected from the first step.
 We propose using different estimating equations for
 $(\alpha,\beta)$, derived by minimizing the asymptotic squared bias
 of  $\widehat{\mu}_{\dr}(\widehat{\alpha},\widehat{\beta})$.
 This estimation strategy is
 not new; see, e.g., \citet{kim2014doubly} for missing data analyses in low-dimensional data;
 however, we demonstrate
 its new role in high-dimensional data to mitigate the possible selection
 error in the first step.
 In essence, our strategy for estimating
 $(\alpha,\beta)$ renders the first order term in the Taylor expansion
 of $\widehat{\mu}_{\dr}(\widehat{\alpha},\widehat{\beta})$ with respect
 to $(\alpha,\beta)$ to be exactly zero, and the remaining terms are negligible
 under regularity conditions.
 This estimating strategy makes
 the doubly robust estimator root-$n$ consistent if either the sampling probability or the outcome model is correctly specified.
 This also enables us to construct a simple and consistent variance estimator allowing for doubly robust
 inferences.
 Importantly, the proposed estimator allows
 model misspecification of either the sampling score or the outcome
 model.
 In the existing high-dimensional causal inference literature, the doubly robust estimators have been shown to be robust to
 selection errors using
 penalization \citep{farrell2015robust} or  approximation errors using machine
 learning  \citep{chernozhukov2018double}.
 However, this double robustness feature requires both nuisance models to be correctly specified.
 We relax this requirement allowing one of the nuisance models to be misspecified.
We clarify that even though the set of variables for estimation may include the variables that are solely related to the sampling score but not the outcome and therefore may harm efficiency of estimating $\mu$ \citep{de2011covariate,patrick2011implications}, it is important to include these variables for $\widehat{\mu}_{\mathrm{dr}}(\widehat{\alpha},\widehat{\beta})$ to achieve consistency in the case when the outcome model is misspecified and the sampling score model is correctly specified; see Section \ref{sec:A-simulation-study}.

The paper proceeds as follows.
Section \ref{sec:Basic-setup} provides the basic setup of the paper.
Section \ref{sec:Methodology} presents the proposed two-step procedure for variable selection and doubly robust estimation of the finite population mean.
Section \ref{sec:Computation} describes the computation algorithm for solving penalized estimating equations.
Section \ref{sec:Asymptotic-results} presents the theoretical properties for variable selection and doubly robust estimation.
Section \ref{sec:A-simulation-study} reports  simulation results that
illustrate the finite-sample performance of the  proposed method.
In Section \ref{sec:Real-data-application}, we
present an application to analyze a non-probability sample collected
by the Pew Research Centre.
We relegate all proofs to the supplementary material.

\section{Basic Setup\label{sec:Basic-setup}}

\subsection{Notation: Two Samples}

Let $\U=\{1,\ldots,N\}$ be the index set of $N$ units for the finite
population, with $N$ being the known population size. The finite
population consists of $\F_{N}=\{(X_{i},Y_{i}):i\in\U\}$. Let the parameter of interest
be the finite population mean $\mu=N^{-1}\sum_{i=1}^{N}Y_{i}$. We consider
two data sources: one from a probability sample, referred to as Sample
A, and the other one from a non-probability sample, referred to as
Sample B. Table \ref{tab:Two-data-sources} illustrates the observed
data structure.
Sample
A consists of observations $\mathcal{O}_{A}=\{(d_{A,i}=\pi_{A,i}^{-1},X_{i}):i\in\A\}$
with sample size $n_{A},$ where $\pi_{A,i}=\pr(i\in\A)$ is known
throughout Sample A, and Sample B consists of observations $\mathcal{O}_{B}=\{(X_{i},Y_{i}):i\in\B\}$
with sample size $n_{B}$. We define $I_{A,i}$ and $I_{B,i}$ to
be the indicators of selection to Sample A and Sample B, respectively.
 Although the non-probability sample contains rich information on $(X,Y)$,
the sampling mechanism is unknown, and therefore we cannot compute the first-order
inclusion probability for Horvitz--Thompson estimation. The naive estimators
without adjusting for the sampling process are subject to selection
biases \citep{meng2018statistical}. On the other hand, although the
probability sample with sampling weights represents the finite population,
it does not observe the study variable of interest.

\begin{table}[t]
	\begin{centering}
		{\scriptsize{}{}{}\caption{\label{tab:Two-data-sources}Two data sources. ``$\protect\surd$''
				and ``?'' indicate observed and unobserved data, respectively. }
		}\centering{}%
		\begin{tabular}{ccccc}
			\hline
			&  & Sample weight  $\pi^{-1}$ & \multicolumn{1}{c}{Covariate $X$ } & Study Variable $Y$\tabularnewline
			\hline
			Probability  & $1$  & $\surd$  & $\surd$  & ? \tabularnewline
			Sample  & $\vdots$  & $\vdots$  & $\vdots$  & $\vdots$ \tabularnewline
			$\mathcal{O}_{A}$  & $n_{A}$  & $\surd$  & $\surd$  & ? \tabularnewline
			\hline
			Non-probability  & $n_{A}+1$  & ?  & $\surd$  & $\surd$ \tabularnewline
			Sample  & $\vdots$  & $\vdots$  & $\vdots$  & $\vdots$ \tabularnewline
			$\mathcal{O}_{B}$  & $n_{A}+n_{B}$  & ?  & $\surd$  & $\surd$ \tabularnewline
			\hline
		\end{tabular}
	Sample A is a probability sample, and Sample B is a non-probability
	sample. 	\par\end{centering}
\end{table}

\subsection{An Identification Assumption}
 Before presenting the proposed methodology for integrating the two data sources, we first discuss the identification assumption.
Let $f(Y\mid X)$ be the conditional distribution of $Y$ given $X$
in the superpopulation model $\zeta$ that generates the finite population.
We make the following primary assumption.

\begin{assumption}\label{asmp:MAR}
	(i) The sampling indicator $I_{B}$ of Sample B and the response variable
	$Y$ is independent given $X$; i.e. $P(I_{B}=1\mid X,Y)=P(I_{B}=1\mid X)$,
	referred to as the sampling score $\pi_{B}(X)$, and (ii)\textcolor{black}{{}
		$\pi_{B}(X)>N^{\gamma-1}\delta_{B}>0$ }for all $X$, where $\gamma\in(2/3,1]$.
	
\end{assumption}

Assumption \ref{asmp:MAR} (i) implies that $E(Y\mid X)=E(Y\mid X,I_{B}=1)$,
denoted by $m(X)$, can be estimated based solely on Sample B.
Assumption
\ref{asmp:MAR} (ii) specifies a lower bound
of $\pi_{B}(X)$ for the technicality in  Section  \ref{sec:Asymptotic-results}.\textcolor{black}{{}
	A standard condition in the literature imposes a strict positivity
	in the sense that $\pi_{B}(X)>\delta_{B}>0$; however, it implies
	that $n_{B}^{-1}=O(N^{-1})$, which may be restrictive in survey sampling. Here,
	we relax this condition and allow $n_{B}^{-1}=O(N^{-\gamma})$, where $\gamma$ can be strictly less than $1$.}

Assumption \ref{asmp:MAR} is a key assumption for identification. Under Assumption \ref{asmp:MAR}, $E(\mu)$ is identifiable based on Sample A by $E\{I_A m(X)\}$ or Sample B by $E\{I_BY/\pi_B(X) \}$.
However, this assumption is not verifiable from the observed data. 
To ensure this assumption holds, researchers often consider many potentially
predictors for the sampling indicator $I_{B}$ or the outcome $Y$, resulting in a rich set of
variables in $X$.

\subsection{Existing Estimators}

In practice, the sampling score function $\pi_{B}(X)$ and the outcome
mean function $m(X)$ are unknown and need to be estimated from the
data. Let $\pi_{B}(X^\T\alpha)$ and $m(X^\T\beta)$ be the posited models
for $\pi_{B}(X)$ and $m(X)$, respectively, where $\alpha$ and $\beta$
are unknown parameters.
Researchers have proposed various estimators for $\mu$ requiring different model assumptions and estimation strategies. We provide
examples below and discuss their properties and limitations.

\begin{example}[Inverse probability of sampling score weighting]Given an estimator $\widehat{\alpha}$, the inverse probability of sampling score weighting estimator is
	\begin{equation}
		\widehat{\mu}_{\ipw}=\widehat{\mu}_{\ipw}(\widehat{\alpha})=\frac{1}{N}\sum_{i=1}^{N}\frac{I_{B,i}}{\pi_{B}(X_{i}^\T\widehat{\alpha})}Y_{i}.\label{eq:psw}
	\end{equation}
\end{example}

There are different approaches to obtain $\widehat{\alpha}$. Following \cite{valliant2011estimating},
one can obtain $\widehat{\alpha}$ by
fitting the sampling score model
based on the blended data $\mathcal{O}_A\cup\mathcal{O}_B=\{(d_{A,i},X_i,I_i=0):i\in\mathcal{A} \}\cup\{(X_i,I_i=1):i\in\mathcal{B} \}$,
weighted by the design weights from Sample A. The resulting estimator $\widehat\alpha$ is valid if the
size of Sample B is relatively small \citep{valliant2011estimating}.
\cite{elliott2017inference}  propose an alternative strategy based on the Bayes rule: $\pi_B(X)\propto P(I_{A}=1\mid X) O_B(X)$, where $O_B(X)=P(I_B=1\mid X)/P(I_B=0\mid X)$ is the odds of selection into Sample B among the blended sample. This approach does not require the  size of Sample B to be small; however, if $X$ does not correspond to the design variables for Sample A, it requires positing an additional model for $P(I_{A}=1\mid X)$. More importantly variable selection based on this approach is not straightforward in the setting with a high-dimensional $X$.
To obtain $\hat{\alpha}$, we use the following estimating equation for  $\alpha$:
\begin{equation}
 \sum_{i=1}^N \left\{  \frac{ I_{B,i} }{ \pi ( X_i^\T \alpha)} - \frac{ I_{A,i}}{ \pi_{A,i} } \right\} h(X_i; \alpha)   = 0,
\label{2-2}
\end{equation}
for some $h(X_i; \alpha)$ such that  (\ref{2-2}) has a unique solution.  \cite{kott2019} advocated using $h(X; \alpha) = X$ and \cite{chen2018doubly}  advocated using $h(X;  \alpha) = \pi(X; \alpha) \cdot X$.
The justification for $\widehat{\mu}_{\ipw}$ relies on the correct
specification of $\pi_{B}(X)$ and the consistency of $\widehat{\alpha}$.
If $\pi_{B}(X_{i}^\T\alpha)$ is misspecified or $\widehat{\alpha}$
is inconsistent, $\widehat{\mu}_{\ipw}$ is biased.

\begin{example}[Outcome regression based on Sample A]The
	outcome regression estimator is
	\begin{equation}
		\widehat{\mu}_{\reg}=\widehat{\mu}_{\reg}(\widehat{\beta})=\frac{1}{N}\sum_{i=1}^{N}I_{A,i}d_{A,i}m(X_{i}^\T\widehat{\beta}),\label{eq:reg}
	\end{equation}
where $\widehat{\beta}$ is obtained by fitting the outcome model based solely on $\mathcal{O}_B=\{(X_i,Y_i):i\in\mathcal{B} \}$  under Assumption \ref{asmp:MAR}.
	
\end{example}

The justification for $\widehat{\mu}_{\reg}$ relies on the correct
specification of {$m(X^T \beta)$} and the consistency of $\widehat{\beta}$.
If $m(X^\T\beta)$ is misspecified or $\widehat{\beta}$ is inconsistent,
$\widehat{\mu}_{\reg}$ can be biased.

\begin{example}[Calibration weighting]The calibration weighting estimator
	is
	\begin{equation}
		\widehat{\mu}_{\Cal}=\widehat{\mu}_{\Cal}=\frac{1}{N}\sum_{i=1}^{N}\omega_{i}I_{B,i}Y_{i},\label{eq:cal}
	\end{equation}
	where $\{\omega_{i}:i\in\mathcal{S}_{B}\}$ satisfies constraint (i)
	$
	\sum_{i\in\mathcal{S}_{B}}\omega_{i}X_{i}=\sum_{i\in\mathcal{S}_{A}}d_{A,i}X_{i},
	$
	or constraint (ii) $
	\sum_{i\in\mathcal{S}_{B}}\omega_{i}m(X_{i};\widehat{\beta})=\sum_{i\in\mathcal{S}_{A}}d_{A,i}m(X_{i};\widehat{\beta})
	$ in a model-assisted approach \citep{ mcconville2017model,chen2018model,chen2019calibrating}.
	
\end{example}

The justification for $\widehat{\mu}_{\Cal}$ subject to constraint (i) relies on the linearity
of the outcome model, i.e., $m(X)=X^{\T}\beta^{*}$ for some $\beta^{*}$,
or the linearity of the inverse probability of sampling weight, i.e., $\pi_{B}(X)^{-1}=X^{\T}\alpha^{*}$
for some $\alpha^{*}$ (\citealp{fuller2009sampling}; Theorem 5.1).
The linearity conditions are unlikely to hold for non-continuous variables. In these cases, $\widehat{\mu}_{\Cal}$ is biased.
The justification for $\widehat{\mu}_{\Cal}$ subject to constraint (ii) relies on
$m(X;\beta)$ being correctly specified in the data integration problem.

\begin{example}[Doubly robust estimator]The doubly robust estimator
	is
	
	\begin{equation}
		\widehat{\mu}_{\dr}=\widehat{\mu}_{\dr}(\widehat{\alpha},\widehat{\beta})=\frac{1}{N}\sum_{i=1}^{N}\left[\frac{I_{B,i}}{\widehat{\pi}_{B}(X_{i}^\T\widehat{\alpha})}\{Y_{i}-m(X_{i};\widehat{\beta})\}+I_{A,i}d_{A,i}m(X_{i};\widehat{\beta})\right].\label{eq:dr}
	\end{equation}
	
\end{example}

The estimator $\widehat{\mu}_{\dr}$ is doubly robust with fixed-dimensional $X$ \citep{chen2018doubly},
in the sense that it achieves the consistency if either $\pi_{B}(X_{i}^\T\alpha)$
or $m(X^\T\beta)$ is correctly specified, but not necessarily both. The double
robustness is attractive; therefore, we shall investigate the potential
of $\widehat{\mu}_{\dr}$ in high-dimensional setup.

\section{Methodology in High-dimensional Data\label{sec:Methodology}}

A major challenge arises in the presence of a large number of covariates,
not all of them are necessary for making inference of the population
mean of the outcome. This necessitates variable selection.
For simplicity of exposition, we introduce the following notation.
For any vector $\alpha\in\R^{p}$, denote the number of nonzero elements in $\alpha$ as $||\alpha||_{0}=\sum_{j=1}^{p}I(\alpha_{j}\neq0)$,
the $L_1$-norm as $||\alpha||_{1}=\sum_{j=1}^{p}|\alpha_{j}|$,
the $L_2$-norm as $||\alpha||_{2}=\sqrt{\sum_{j=1}^{p}\alpha_{j}^{2}}$, and
the $L_\infty$-norm as $||\alpha||_{\infty}=\max_{j=1}^{p}|\alpha_{j}|$.
For any $\J\subseteq\{1,\ldots,p\}$,
let $\alpha_{\J}$ be the sub-vector of $\alpha$ formed by elements
of $\alpha$ whose indexes are in $\J$. Let $\J^{c}$ be the complement
of $\J$. For any $\J_{1},\J_{2}\subseteq\{1,\ldots,p\}$ and
matrix $\Sigma\in\R^{p\times p}$, let $\Sigma_{\J_{1},\J_{2}}$ be
the sub-matrix of $\Sigma$ formed by rows in $\J_{1}$ and columns
in $\J_{2}$.
Following the literature on variable selection, we can first
standardize the covariates so that approximately they have variances
equal to one to stabilize the variable selection procedure.
We make the following modeling assumptions.

\begin{assumption}[Sampling score model]\label{assump:psm}
The sampling mechanism of Sample B, $\pi_{B}(X)$, follows
	a logistic regression model $\pi_{B}(X^\T \alpha)$; i.e., $\logit\left\{ \pi_{B}(X^\T \alpha)\right\} =X^{\T}\alpha$
	for $\alpha\in\R^{p}$.

\end{assumption}

\begin{assumption}[Outcome model]
	The outcome mean function  $m(X)$ follows a generalized linear regression model; i.e., $m(X)=m(X^{\T}\beta)$ for $\beta\in\R^{p}$, where  $m(\cdot)$ is a link function by an abuse the notation.	
\end{assumption}

Define $\alpha^{*}$ to be the $p$-dimensional
parameter that minimizes the Kullback-Leibler divergence
\[
\alpha^{*}=\arg\min_{\alpha\in\R^{p}}E\left[\pi_{B}(X) \log\frac{\pi_{B}(X)}{\pi_{B}(X^\T\alpha)}+\{1- \pi_{B}(X) \}\log\frac{1-\pi_{B}(X)}{1-\pi_{B}(X^\T\alpha)}\right],
\]
and $\beta^{*}=\arg\min_{\beta}E\left[\{Y-m(X^\T\beta)\}^{2}\right].$

In Assumption \ref{assump:psm}, we adopt the logistic regression model for the sampling score following most of the empirical literature; but our framework can be extended to the case with other parameter models such as the probit model.
The models $\pi_{B}(X^\T\alpha)$ and $m(X^\T\beta)$  are working models
and they may be misspecified. If the sampling score model is correctly
specified, $\pi_{B}(X)=\pi_{B}(X^\T\alpha^{*})$. If the outcome model
is correctly specified, $m(X)=m(X^\T\beta^{*})$.


The proposed procedure consists of two steps: the first step selects important variables in the sampling score model and the outcome model, and the second step focuses on doubly robust estimation of the population mean.

In the first step, we propose solving penalized estimating equations for variable selection.
Using (\ref{2-2}) with $h(X; \alpha)=X$, we define the estimating function for $\alpha$ as
\[
U_{1}(\alpha)=\frac{1}{N}\sum_{i=1}^{N}\left\{ \frac{I_{B,i}}{\pi_{B}(X_{i}^{\T}\alpha)}-\frac{I_{A,i}}{\pi_{A,i}}\right\} X_{i}.
\]
To select important variables in $m(X^\T\beta)$, under Assumption \ref{asmp:MAR},
we have $E(Y\mid X)=E(Y\mid X,I_{B}=1)$. Therefore, we define the
estimating function for $\beta$ as
\[
U_{2}(\beta)=\frac{1}{N}\sum_{i=1}^{N}I_{B,i}\left\{ Y_{i}-m(X_{i}^{\T}\beta)\right\} X_{i}.
\]
Let $U(\theta)=(U_{1}(\alpha)^{\T},U_{2}(\beta)^{\T})^{\T}$ be the
joint estimating function for $\theta=(\alpha^{\T},\beta^{\T})^{\T}$.
When $p$ is large, following \citet{johnson2008penalized}, we consider
solving the penalized estimating function
\begin{equation}
	U^{\p}(\alpha,\beta)=U(\alpha,\beta)-\left(\begin{array}{c}
		q_{\lambda_{\alpha}}(|\alpha|)\sign(\alpha)\\
		q_{\lambda_{\beta}}(|\beta|)\sign(\beta)
	\end{array}\right),\label{eq:adaptive lasso}
\end{equation}
for $(\alpha,\beta)$, where $q_{\lambda_{\alpha}}(\alpha)=\{q_{\lambda_{\alpha}}(|\alpha_{0}|),\ldots,q_{\lambda_{\alpha}}(|\alpha_{p}|)\}^{\T}$
and $q_{\lambda_{\beta}}(\beta)=\{q_{\lambda_{\beta}}(|\beta_{0}|),\ldots,q_{\lambda_{\beta}}(|\beta_{p}|)\}^{\T}$
are some continuous functions, $q_{\lambda_{\alpha}}(|\alpha|)\sign(\alpha)$
is the element-wise product of $q_{\lambda_{\alpha}}(\alpha)$ and
$\sign(\alpha),$ and $q_{\lambda_{\beta}}(|\beta|)\sign(\beta)$
is the element-wise product of $q_{\lambda_{\beta}}(\beta)$ and $\sign(\beta)$.
We let $q_{\lambda}(x)=\de p_{\lambda}(x)/\de x$, where $p_{\lambda}(x)$
is some penalization function.
Although the same discussion applies to different non-concave penalty functions, we specify $p_{\lambda}(x)$ to be
a folded-concave smoothly clipped absolute deviation (SCAD) penalty function \citep{fan2011nonconcave}.
Accordingly,  we have
\begin{equation}
	q_{\lambda}(|\theta|)=\lambda\left\{ I(|\theta|<\lambda)+\frac{(a\lambda-|\theta|)_{+}}{(a-1)\lambda}I(|\theta|\geq\lambda)\right\} ,\label{eq:q_lambda}
\end{equation}
for $a>0$, where $(\cdot)_+$ is the truncated linear function; i.e., if $x\geq0$, $(x)_+=x$, and if $x<0$,  $(x)_+=0$.
We use $a=3.7$.
 \citet{fan2001variable} demonstrate that with $a=3.7$, the SCAD selector has a good performance based on simulation. Thereafter, this choice has become standard in the literature and become a default choice in many softwares such as ``ncvreg" in R.
We select the variables if the corresponding estimates of
(\ref{eq:adaptive lasso}) are nonzero in either the sampling
score or the outcome model,
indexed by $\C$.

\begin{remark}
To help understand the penalized estimating equation, we discuss two
scenarios. If $|\alpha_{j}|$ is large, then $q_{\lambda_{\alpha}}(|\alpha_{j}|)$
is zero, and therefore $U_{1,j}(\alpha)$ is not penalized. Whereas,
if $|\alpha_{j}|$ is small but nonzero, then $q_{\lambda_{\alpha}}(|\alpha_{j}|)$
is large, and therefore $U_{1,j}(\alpha)$ is penalized with a penalty
term. The penalty term then forces $\widehat{\alpha}_{j}$ to be zero and
excludes the $j$th element in $X$ from the final selected set of
variables. The same discussion applies to $U_{2}(\beta)$ and $q_{\lambda_{\beta}}(|\beta|)$.
\end{remark}


In the second step,  we consider the estimator of the population mean
$\widehat{\mu}_\dr(\widehat{\alpha},\widehat{\beta})$ in (\ref{eq:dr})
 with $(\widehat{\alpha},\widehat{\beta})$  re-estimated based on $X_{\C}$.
As we will show in Section \ref{sec:Asymptotic-results}, $\C$ contains
the true important variables in either the sampling score model or the outcome model with probability approaching one (the
oracle property). Therefore, if either the sampling
score model or  the outcome model is correctly specified, the asymptotic bias of $\widehat{\mu}_{\dr}(\alpha^{*},\beta^{*})$
is zero; however, if both models are misspecified, the asymptotic
bias of $\widehat{\mu}_{\dr}(\alpha^{*},\beta^{*})$ is
\begin{multline*}
	\text{a.bias}(\alpha^{*},\beta^{*})  =  E\left\{ \widehat{\mu}_{\dr}(\alpha^{*},\beta^{*})-\mu\right\} \\
	=  E\left[\frac{1}{N}\sum_{i=1}^{N}\left\{ \frac{I_{B,i}}{\pi_{B}(X_{i}^\T\alpha^{*})}-1\right\} \{Y_{i}-m(X_{i}^\T\beta^{*})\}\right]
	  +E\left\{ \frac{1}{N}\sum_{i=1}^{N}\left(I_{A,i}d_{A,i}-1\right)m(X_{i}^\T\beta^{*})\right\} .
\end{multline*}
In order to minimize $\{\text{a.bias}(\alpha,\beta)\}^{2}$,
we consider the estimating function
	\begin{equation}
\frac{\partial \{\text{a.bias}(\alpha^{},\beta^{})\}^{2} }{\partial (\alpha_\C^\T,\beta_\C^\T)^\T}
= 2 \times \text{a.bias}(\alpha^{},\beta^{}) \times
\left(\begin{array}{c}
I_{B}\left\{ \frac{1}{\pi_{B}(X^\T\alpha)}-1\right\} \{Y-m(X^{\T}\beta)\}X_{\C}\\
\left\{ \frac{I_{B}}{\pi_{B}(X^\T\alpha)}-d_{A}I_{A}\right\} \partial m(X^{\T}\beta)/\partial\beta_{\C}
\end{array}\right)\label{eq:jee0}
\end{equation}
	 and the corresponding empirical estimating function
\begin{equation}
	J(\alpha,\beta)=\left(\begin{array}{c}
		J_{1}(\alpha,\beta)\\
		J_{2}(\alpha,\beta)
	\end{array}\right)=\left(\begin{array}{c}
		\frac{1}{N}\sum_{i=1}^{N}I_{B,i}\left\{ \frac{1}{\pi_{B}(X_{i}^\T\alpha)}-1\right\} \{Y_{i}-m(X_{i}^{\T}\beta)\}X_{i\C}\\
		\frac{1}{N}\sum_{i=1}^{N}\left\{ \frac{I_{B,i}}{\pi_{B}(X_{i}^\T\alpha)}-d_{A,i}I_{A,i}\right\} \partial m(X_{i}^{\T}\beta)/\partial\beta_{\C}
	\end{array}\right)\label{eq:jee}
\end{equation}
for estimating $(\alpha,\beta)$, constrained on $\{(\alpha^\T,\beta^\T)^\T\in\R^{2p}:\alpha_{\C^{c}}=0,\beta_{\C^{c}}=0\}$.
Equation (\ref{eq:jee}) is doubly robust in the sense that $J(\alpha^*,\beta^*)$ is unbiased if either $\pi(X^{\T}\alpha)$ or $m(X^{\T}\beta)$ is correctly specified, not necessarily both \citep{kim2014doubly}.

\begin{remark}
The two steps use different estimating functions (\ref{eq:adaptive lasso})
and (\ref{eq:jee}), respectively, for selection and estimation with the following
advantages. First, (\ref{eq:adaptive lasso}) separates the selection
for $\alpha$ and $\beta$ in $U_{1}(\alpha)$ and $U_{2}(\beta)$,
so it stabilizes the selection procedure if either the
sampling score model or the outcome model is misspecified.
{Second, using (\ref{eq:jee}) for estimation leads to an attractive feature for inference about $\mu$.
We clarify that although the joint estimating function (\ref{eq:jee}) is motivated by minimizing the asymptotic bias $\text{a.bias}(\alpha^*,\beta^*)$ if both nuisance models are misspecified,
we do not expect that the proposed estimator for $\mu$ is unbiased in this case.
Instead, we show the advantage of (\ref{eq:jee}) in the case if either
the sampling probability or the outcome model is correctly specified in high-dimensional data.
It is well-known that post-selection inference is notoriously difficult even when both models are correctly specified because the estimation step is based on a random set of variables being selected.
We show that our estimating strategy based on (\ref{eq:jee})  mitigates the possible first-step selection error
and renders
$\widehat{\mu}_\dr(\widehat{\alpha},\widehat{\beta})$
 root-$n$ consistent if either
the sampling probability or the outcome model is correctly specified in high-dimensional data. Heuristically this is achieved because the first Taylor expansion term is set to be zero due to (\ref{eq:jee0}).
We relegate the details to Section \ref{sec:Asymptotic-results}.
}
\end{remark}

In summary, our two-step procedure for variable selection and estimation
is as follows.
\begin{list}{label}{spacing}
		\item[{Step$\ 1.$}] To facilitate joint selection of variables for the
	sampling score and outcome, solve the penalized joint
	estimating equations $U^{\p}(\alpha,\beta)=0$ in (\ref{eq:adaptive lasso}),
	denoted by $(\widetilde{\alpha},\widetilde{\beta})$. Let $\widehat{\M}_{\alpha}=\{j:\widetilde{\alpha}_{j}\neq0\}$
	and $\widehat{\M}_{\beta}=\{j:\widetilde{\beta}_{j}\neq0\}$.
	\item[{Step$\ 2.$}] Let the set of variables for estimation be $\C=\widehat{\M}_{\alpha}\cup\widehat{\M}_{\beta}$.
	Obtain the proposed estimator as
	\begin{equation}
	\widehat{\mu}_{\pee}=\widehat{\mu}_{\pee}(\widehat{\alpha},\widehat{\beta})=\frac{1}{N}\sum_{i=1}^{N}\left\{ I_{B,i}\frac{Y_{i}-m(X_{i}^{\T}\widehat{\beta})}{\pi_{B}(X_{i}^\T\widehat{\alpha})}+I_{A,i}d_{A,i}m(X_{i}^{\T}\widehat{\beta})\right\} ,\label{eq:pee}
	\end{equation}
	where $\widehat{\alpha}$ and $\widehat{\beta}$ are obtained by solving
	the joint estimating equations (\ref{eq:jee}) for $\alpha$ and $\beta$
	with $\alpha_{\C^{c}}=0$ and $\beta_{\C^{c}}=0$.
\end{list}

\begin{remark}\label{rmk:joint}
	Variable selection circumvents the instability or infeasibility of direct estimation of $(\alpha,\beta)$ with high-dimensional $X$.
	Moreover, in Step 2 for estimation, we consider a union of covariates $X_{\C}$, where $\C=\widehat{\M}_{\alpha}\cup\widehat{\M}_{\beta}$.
	{It is worth comparing this choice with two other common choices in the literature.
First, one considers separate sets of variables for the two models; i.e., the sampling score is fitted based on  $\widehat{\M}_{\alpha}$, and the outcome model is fitted based on
$\widehat{\M}_{\beta}$. However, we note that in the joint estimating equation (\ref{eq:jee}), $J_1(\alpha,\beta)$ and $J_2(\alpha,\beta)$ should have the same dimension, otherwise, it is possible that (\ref{eq:jee}) does not guarantee that there exists a solution. This is obvious if one considers a linear outcome model. Moreover, \citet{brookhart2006variable} and \citet{shortreed2017outcome} show
that including variables that are related to the outcome in the propensity score model will increase the precision of the estimated average treatment effect without increasing
bias.  This implies that an efficient variable selection and estimation
method should take into account both sampling-covariate and outcome-covariate
relationships.
As a result,  $\widehat{\mu}_\dr(\widehat{\alpha},\widehat{\beta})$ may have a better performance than the oracle estimator which uses the true important variables
in the sampling score and the outcome model. This is particularly
true when one of the models is misspecified.
Our simulation study
in Section \ref{sec:A-simulation-study} demonstrates that $\widehat{\mu}_{\dr}$
with variable selection has a similar performance as the orcale estimator
for the continuous outcome and outperforms the oracle estimator for
the binary outcome.
Second, many authors have suggested that including predictors that are solely related to the sampling score but not the outcome may harm efficiency \citep{de2011covariate,patrick2011implications}.
However, this strategy is effective when both the sampling score and outcome models are correctly specified. When the sampling score model is correctly specified but the outcome model is misspecified, restricting the variables to be the outcome predictors may render the sampling score ``misspecified" by using the wrong set of variables. The simulation study suggests that $\widehat{\mu}_{\mathrm{p}-dr}$ restricted to the set of variables in $\widehat{\mathcal{M}}_\beta$ is not doubly robust.
}

\end{remark}

\section{Computation\label{sec:Computation} }

In this section, we discuss the computation for solving the penalized estimating function (\ref{eq:adaptive lasso}).
Following \citet{johnson2008penalized}, we use an iterative algorithm that combines
the Newton--Raphson algorithm for solving estimating equation and the
minorization-maximization algorithm for non-convex penalty of \citet{hunter2005variable}.

First, by the minorization-maximization algorithm, the penalized estimator
$\widetilde{\theta}=(\widetilde{\alpha},\widetilde{\beta})$ solving (\ref{eq:adaptive lasso})
satisfies
\begin{equation}
U^{\p}(\widetilde{\theta})=U(\widetilde{\theta})-\left(\begin{array}{c}
q_{\lambda_{\widetilde\alpha}}(|\widetilde\alpha|)\sign(\widetilde\alpha)\frac{|\widetilde{\alpha}|}{\epsilon+|\widetilde{\alpha}|}\\
q_{\lambda_{\widetilde\beta}}(|\widetilde\beta|)\sign(\widetilde\beta)\frac{|\widetilde{\beta}|}{\epsilon+|\widetilde{\beta}|}
\end{array}\right)=0,\label{eq:(5)}
\end{equation}
for $\epsilon$ is a predefined small number. In our implementation,
we choose $\epsilon$ to be $10^{-6}$.

Second, we solve (\ref{eq:(5)}) by the Newton-Raphson algorithm.
It may be challenging to implement the Newton-Raphson algorithm directly,
because it involves inverting a large matrix. For ease and stability in those cases, we can use a coordinate
decent algorithm \citep{friedman2007pathwise} by cycling through
and updating each of the coordinates.

Following most of the empirical literature, we assume that $\pi_{B}(X^\T\alpha)$
follows a logistic regression model. Define $m^{(k)}(t)=\de^{k}m(t)/\de^{k}t$
for $k\geq1$. We denote
\begin{eqnarray}
\nabla(\theta)=\partial U(\theta)/\partial\theta^{\T} & = & \text{diag}\{\partial U_{1}(\alpha)/\partial\alpha^{\T},\partial U_{2}(\beta)/\partial\beta^{\T}\},\label{eq:Delta}\\
\frac{\partial U_{1}(\alpha)}{\partial\alpha^{\T}} & = & -\frac{1}{N}\sum_{i=1}^{N}I_{B,i}\frac{1-\pi_{B}(X_{i}^\T\alpha)}{\pi_{B}(X_{i}^\T\alpha)}X_{i}X_{i}^{\T},\nonumber \\
\frac{\partial U_{2}(\beta)}{\partial\beta^{\T}} & = & -\frac{1}{N}\sum_{i=1}^{N}I_{B,i}m^{(1)}(X_{i}^{\T}\beta)^{2}X_{i}X_{i}^{\T},\nonumber
\end{eqnarray}
and
\begin{eqnarray*}
	E(\theta) & = & \left(\begin{array}{ccc}
		q_{\lambda_{1}}(|\theta_{1}|) & \cdots & 0\\
		\vdots & \ddots & \vdots\\
		0 & \cdots & q_{\lambda_{2p}}(|\theta_{2p}|)
	\end{array}\right).
\end{eqnarray*}
Let $\theta$ start at an initial value $\widetilde{\theta}^{[0]}$.
With the other coordinates fixed, the $k$th Newton-Raphson update
for $\theta_{j}$ ($j=1,\ldots,2p$), the $j$th element of $\theta$,
is
\begin{equation}
\widetilde{\theta}_{j}^{[k]}=\widetilde{\theta}_{j}^{[k-1]}+\left\{ \nabla_{jj}(\widetilde{\theta}^{[k-1]})+N\cdot E_{jj}(\widetilde{\theta}^{[k-1]})\right\} ^{-1}\left\{ U_{j}(\widetilde{\theta}^{[k-1]})-N\cdot E_{jj}(\widetilde{\theta}^{[k-1]})\widetilde{\theta}_{j}^{[k-1]}\right\} ,\label{eq:(4.2)}
\end{equation}
where $\nabla_{jj}(\theta)$ and $E_{jj}(\theta)$ are the $j$th
diagonal elements in $\nabla(\theta)$ and $E(\theta)$, respectively.
The procedure cycles through all the $2p$ elements of $\theta$ and
is repeated until convergence.

We use $K$-fold cross-validation to select the tuning parameter
$(\lambda_{\alpha}, \lambda_{\beta})$. To be specific, we partition
both samples into approximately $K$ equal sized subsets and pair
subsets of Sample A and subsets of Sample B randomly. Of the $K$
pairs, we retain one single pair as the validation data and the remaining
$K-1$ pairs as the training data. We fit the models based on the training
data and estimate the loss function based on the validation data.
We repeat the process $K$ times, with each of the $K$ pairs used
exactly once as the validation data. Finally, we aggregate the $K$
estimated loss function. We select the tuning parameter as the one
that minimizes the aggregated loss function over a pre-specified grid.

Because the weighting estimator uses the sampling score $\pi_{B}(X_{})$
to calibrate the distribution of $X_{\C}$ between Sample B and the
target population, we use the following loss function for selecting
$\lambda_{\alpha}$:
\[
\mathrm{Loss}(\lambda_{\alpha})=\sum_{j=1}^{p}\left[\sum_{i=1}^{N}\left\{ \frac{I_{B,i}}{\pi_{B}\{X_{i}^\T\widehat{\alpha}(\lambda_{\alpha})\}}-\frac{I_{A,i}}{\pi_{A,i}}\right\} X_{i,j}\right]^{2},
\]
where $\widetilde{\alpha}(\lambda_{\alpha})$ is the penalized estimator
$\widetilde{\alpha}$ with the tuning parameter $\lambda_{\alpha}$.
We use the prediction error loss function for selecting $\lambda_{\beta}$:
\[
\mathrm{Loss}(\lambda_{\beta})=\sum_{i=1}^{N}I_{B,i}\left[Y_{i}-m\{X_{i}^{\T}\widehat{\beta}(\lambda_{\beta})\}\right]^{2},
\]
where $\widetilde{\beta}(\lambda_{\beta})$ is the penalized estimator
$\widetilde{\beta}$ with the tuning parameter $\lambda_{\beta}$.

\section{Asymptotic Results for Variable Selection and Estimation\label{sec:Asymptotic-results} }

We establish the asymptotic properties for the proposed double variable
selection and doubly robust estimation.
We can establish  theoretical
results for general sampling mechanisms for Sample A requiring specific
regularity conditions.
In this section, for technical convenience,
we assume that  Sample A is collected by simple random
sampling or Poisson sampling with the following regularity conditions.

\begin{assumption}\label{asmp:sampling}
	For all $1\leq i\leq N$, $\pi_{A,i}\geq N^{\gamma-1}\delta_{A}>0$,
	where $\gamma\in(2/3,1]$.
	
\end{assumption}

Similar to Assumption \ref{asmp:MAR} (ii), we relax
	the strict positivity on $\pi_{A,i}$  and render $n_{A}=O(N^{\gamma})$ for $\gamma$ possibly strictly less than $1$.
Let $n=\min(n_{A},n_{B})$, which is $O(N^{\gamma})$ under Assumptions \ref{asmp:MAR} and \ref{asmp:sampling}.

Let the support of model parameters be
\[
\M_{\alpha}=\{1\leq j\leq p:\alpha_{j}^{*}\neq0\},\ \ \M_{\beta}=\{1\leq j\leq p:\beta_{j}^{*}\neq0\},\ \ \M_{\theta}=\M_{\alpha}\cup\{p+\M_{\beta}\}.
\]
 Define $s_{\alpha}=||\alpha^{*}||_{0}$,
$s_{\beta}=||\beta^{*}||_{0}$ , $s_{\theta}=s_{\alpha}+s_{\beta}$,
and $\lambda_{\theta}=\min(\lambda_{\alpha},\lambda_{\beta})$.

\begin{assumption}\label{assump:regularity } The following regularity
	conditions hold.
	\begin{description}
		\item [{(A1)}] The parameter $\theta$ belongs to a compact subset in $\R^{2p}$,
		and $\theta^{*}$ lies in the interior of the compact subset.
		\item [{(A2)}] $\{ X_{i}:  i\in \mathcal U\}$  are fixed and uniformly bounded.
		\item [{(A3)}] There exist constants $c_{1}$ and $c_{2}$ such that
		\[
		0<c_{1}\leq\lambda_{\min}\left(\frac{1}{N}\sum_{i=1}^{N}X_{i}^{\T}X_{i}\right)\leq\lambda_{\max}\left(\frac{1}{N}\sum_{i=1}^{N}X_{i}^{\T}X_{i}\right)\leq c_{2}<\infty,
		\]
		where $\lambda_{\min}(\cdot)$ and $\lambda_{\max}(\cdot)$ are the
		minimum and the maximum eigenvalue of a matrix, respectively.
		\item [{(A4)}] Let $\epsilon_{i}(\beta)=Y_{i}-m(X_{i}^{\T}\beta)$ be the
		$i$th residual. There exists a constant $c_{3}$ such that $E\{|\epsilon_{i}(\beta^{*})|^{2+\delta}\}\leq c_{3}$
		for all $1\leq i\leq N$ and some $\delta>0$. There exist constants
		$c_{4}$ and $c_{5}$ such that $E[\exp\{c_{4}|\epsilon_{i}(\beta^{*})|\}\mid X_{i}]\leq c_{5}$
		for all $1\leq i\leq N$.
		\item [{(A5)}] $m^{(1)}(X_{i}^{\T}\beta)$, $m^{(2)}(X_{i}^{\T}\beta)$,
		and $m^{(3)}(X_{i}^{\T}\beta)$ are uniformly bounded away from 
		$\infty$ on
		$\N_{\theta,\tau}  = \{ \theta\in\R^{2p}:||\theta_{\M_{\theta}}-\theta_{\M_{\theta}}^{*}||\leq\tau\sqrt{s_{\theta}/n},\theta_{\M_{\theta}^{c}}=0\}
		$ for some $\tau>0$.
		\item [{(A6)}] $\min_{j\in\M_{\alpha}}|\alpha_{j}^{*}|/\lambda_{\alpha}\rightarrow\infty$
		and $\min_{k\in\M_{\beta}}|\beta_{k}^{*}|/\lambda_{\beta}\rightarrow\infty,$
		as $n\rightarrow\infty$.
		\item [{(A7)}] $s_{\theta}=o(n^{1/3})$, $\lambda_{\alpha},\lambda_{\beta}\rightarrow0$,
		$(\log n)^{2}=o(n\lambda_{\theta}^{2})$, $\log(p)=o\left\{ n\lambda_{\theta}^{2}/\left(\log n\right){}^{2}\right\} $,
		$ps_{\theta}^{4}(\log n)^{6}=o(n^{3}\lambda_{\theta}^{2})$, $ps_{\theta}^{4}(\log n)^{8}=o(n^{4}\lambda_{\theta}^{4})$,
		as $n\rightarrow\infty$.
	\end{description}
\end{assumption}

These assumptions are typical in the penalization literature.\textcolor{black}{{}
	(A2) specifies a fixed design which is well suited under the finite
	population inference framework. }(A4) holds for Gaussian distribution,
sub-Gaussian distribution, and so on. (A5) holds for common models.
(A7) specifies the restrictions on the dimension of covariates $p$ and
the dimension of the true nonzero coefficients $s_{\theta}$. To
gain insight, when the true model size $s_{\theta}$ is fixed, (A7)
holds for $p=O(n)$, i.e., $p$ can be the same size as $n$.

We establish the asymptotic properties of the penalized estimating
equation procedure.

\begin{theorem}\label{Thm:1 selection consistency}
	
	Under Assumptions \ref{asmp:MAR}\textendash \ref{assump:regularity },
	there exists an approximate penalized solution $\widetilde{\theta}$,
	which satisfies the selection consistency properties:
	\begin{eqnarray}
		P(|U_{j}^{\p}(\widetilde{\theta})|=0,j\in\M_{\theta}) & \rightarrow & 1,\label{eq:(6)}\\
		P\left(|U_{j}^{\p}(\widetilde{\theta})|\leq\frac{\lambda_{\theta}}{\log n},j\in\M_{\theta}^{c}\right) & \rightarrow & 1,\label{eq:(7)}\\
		P\left(\widetilde{\theta}_{\M_{\theta}^{c}}=0\right) & \rightarrow & 1,\label{eq:(2)}
	\end{eqnarray}
	and
	\begin{equation}
		\widetilde{\theta}_{\M_{\theta}}-\theta_{\M_{\theta}}^{*}=O_{P}(\sqrt{s_{\theta}/n}),\label{eq:(3)}
	\end{equation}
	as $n\rightarrow\infty.$
	
\end{theorem}

Results (\ref{eq:(6)}) and (\ref{eq:(7)}) imply that $U(\widetilde{\theta})=o_{P}\left(\lambda_{\theta}/\log n\right)$.
Results (\ref{eq:(2)}) and (\ref{eq:(3)}) imply that with probability
approaching to one, the penalized estimating equation procedure would
not over-select irrelevant variables and estimate the true nonzero
coefficients at the $\sqrt{s_{\theta}/n}$ convergence rate, which
is the so-called oracle property of variable selection.

\begin{remark}	
{It is worth discussing the relationship of Theorem \ref{Thm:1 selection consistency} to  existing variable selection methods in the survey literature.
Based on a single probability sample source,
\cite{mcconville2017model} propose a model-assisted survey regression estimator of finite-population totals using the LASSO to improve the efficiency. \cite{chen2018model} and \cite{chen2019calibrating} propose model-assisted calibration estimators using the LASSO based on nonprobability samples integrating with auxiliary known totals or probability samples, respectively. However, their methods require the working outcome model to include sufficient population information and therefore are not doubly robust.    	
To the best of our knowledge, our paper is the first to propose doubly robust inference of finite population means after variable selection.}
\end{remark}

We now establish the asymptotic properties of $\widehat{\mu}_{\pee}(\widehat{\alpha},\widehat{\beta})$.
Define a sequence of events $\mathcal{D}_{n}=\{\M_{\theta}\subset\C\}$,
where we emphasize that $\mathcal{D}_{n}$ depends on $n$ although
we suppress the dependence of $\M_{\theta}$ and $\C$  on $n$. Following
the same argument for (\ref{eq:(3)}), given the event $\mathcal{D}_{n}$,
we have $\{(\widehat{\alpha}-\alpha^{*})^{\T},(\widehat{\beta}-\beta^{*})^{\T}\}=O_{p}(\sqrt{s_{\theta}/n}).$
Combining with $P(\mathcal{D}_{n})\rightarrow1$, we have
\begin{equation}
	\{(\widehat{\alpha}-\alpha^{*})^{\T},(\widehat{\beta}-\beta^{*})^{\T}\}=O_{p}(\sqrt{s_{\theta}/n}).\label{eq:hat alpha}
\end{equation}
By Taylor expansion,
\begin{eqnarray}
	n^{1/2}\left\{ \widehat{\mu}_{\pee}(\widehat{\alpha},\widehat{\beta})-\mu\right\}  & = & n^{1/2}\left\{ \widehat{\mu}_{\pee}(\alpha^{*},\beta^{*})-\mu\right\} +n^{1/2}\left\{ \frac{\widehat{\mu}_{\pee}(\widehat{\alpha},\widehat{\beta})}{\partial(\alpha^{\T},\beta^{\T})}\right\} \left(\begin{array}{c}
		\widehat{\alpha}-\alpha^{*}\\
		\widehat{\beta}-\beta^{*}
	\end{array}\right)\nonumber \\
	&  & +O_{P}\left\{ n^{1/2}\left|\left|\left(\begin{array}{c}
		\widehat{\alpha}-\alpha^{*}\\
		\widehat{\beta}-\beta^{*}
	\end{array}\right)\right|\right|_{2}^{2}\right\} \nonumber \\
		& = & n^{1/2}\left\{ \widehat{\mu}_{\pee}(\alpha^{*},\beta^{*})-\mu\right\} +O_{P}\left\{ n^{1/2}\left|\left|\left(\begin{array}{c}
	\widehat{\alpha}-\alpha^{*}\\
	\widehat{\beta}-\beta^{*}
	\end{array}\right)\right|\right|_{2}^{2}\right\} \label{eq:pee-a}\\
	& = & n^{1/2}\left\{ \widehat{\mu}_{\pee}(\alpha^{*},\beta^{*})-\mu\right\} + o_p(1),\label{eq:pee-a1}
\end{eqnarray}
where $\widehat{\mu}_{\pee}(\alpha,\beta)$ is defined in (\ref{eq:pee}).
Equation (\ref{eq:pee-a}) follows because we solve (\ref{eq:jee})
for $(\alpha,\beta)$. Equation (\ref{eq:pee-a1}) follows because
of (\ref{eq:hat alpha}) and Assumption \ref{assump:regularity }
(A7).  As a result, the way for estimating $(\alpha^{*},\beta^{*})$
leads to the asymptotic equivalence between $\widehat{\mu}_{\pee}(\widehat\alpha,\widehat\beta)$ and $\widehat{\mu}_{\pee}(\alpha^{*},\beta^{*})$.

Moreover, we show that $\widehat{\mu}_{\pee}(\alpha^{*},\beta^{*})$ is asymptotically unbiased for $\mu$ under the double robustness condition. We note that
\begin{equation}
	n^{1/2}E[\left\{ \widehat{\mu}_{\pee}(\alpha^{*},\beta^{*})-\mu\right\}]\nonumber
	= \frac{n^{1/2}}{N}\sum_{i=1}^{N}E\left\{ \frac{I_{B,i}}{\pi_{B}(X_{i}^\T\alpha^{*})}-1\mid X_{i}\right\} E\left\{ Y_{i}-m(X_{i}^{\T}\beta^{*})\mid X_{i}\right\} .\label{eq:dr-1}
\end{equation}
If $\pi_{B}(X^\T\alpha)$ is correctly specified, then $\pi_{B}(X^\T\alpha^{*})=\pi_{B}(X)$
and therefore (\ref{eq:dr-1}) is zero; if $m(X_{i}^{\T}\beta)$ is
correctly specified, then $m(X_{i}^{\T}\beta^{*})=m(X_{i})$ and therefore
(\ref{eq:dr-1}) is zero.

{Following the variance decomposition strategy of \cite{shao1999variance},}
the asymptotic variance of the linearized term is
\begin{multline*}
	V\left[n^{1/2}\left\{ \widehat{\mu}_{\pee}(\alpha^{*},\beta^{*})-\mu\right\} \right]=n^{1/2}E\left[V\left\{ \widehat{\mu}_{\pee}(\alpha^{*},\beta^{*})-\mu\mid I_{B},X,Y\right\} \right]\\
	+n^{1/2}V\left[E\left\{ \widehat{\mu}_{\pee}(\alpha^{*},\beta^{*})-\mu\mid I_{B},X,Y\right\} \right]:=V_{1}+V_{2},
\end{multline*}
where the conditional distribution in $E(\cdot\mid I_{B},X,Y)$ and
$V(\cdot\mid I_{B},X,Y)$ is the sampling distribution for Sample
A. The first term $V_{1}$ is the sampling variance of the Horvitz--Thompson
estimator. Thus,
\begin{equation}
	V_{1}=E\left\{ \frac{n}{N^{2}}\sum_{i=1}^{N}\sum_{j=1}^{N}(\pi_{A,ij}-\pi_{A,i}\pi_{A,j})\frac{m(X_{i}^{\T}\beta^{*})}{\pi_{A,i}}\frac{m(X_{j}^{\T}\beta^{*})}{\pi_{A,j}}\right\} .\label{eq:V1}
\end{equation}
For the second term $V_{2}$, note that
\[
E\left\{ \widehat{\mu}_{\pee}(\alpha^{*},\beta^{*})-\mu\mid I_{B},X,Y\right\} =\frac{1}{N}\sum_{i=1}^{N}\left\{ \frac{I_{B,i}}{\pi_{B,i}(X_{i}^\T\alpha^{*})}-1\right\} \left\{ Y_{i}-m(X_{i}^{\T}\beta^{*})\right\} .
\]
Thus,
\begin{equation}
	V_{2}=\frac{n}{N^{2}}\sum_{i=1}^{N}E\left[\left\{ \frac{I_{B,i}}{\pi_{B,i}(X_{i}^\T\alpha^{*})}-1\right\} ^{2}\left\{ Y_{i}-m(X_{i}^{\T}\beta^{*})\right\} ^{2}\right].\label{eq:V2}
\end{equation}

Theorem \ref{Thm:dr} below summarizes the asymptotic properties of $\widehat{\mu}_{\pee}$.

\begin{theorem}\label{Thm:dr}
	
	Under Assumptions \ref{asmp:MAR}\textendash \ref{assump:regularity },
	if either $\pi_{B}(X^\T\alpha)$ or $m(X^\T\beta)$ is correctly specified,
	\[
	n^{1/2}\left\{ \widehat{\mu}_{\pee}(\widehat{\alpha},\widehat{\beta})-\mu\right\} \rightarrow\N\left(0,V\right),
	\]
	as $n\rightarrow\infty$, where $V=\lim_{n\rightarrow\infty}(V_{1}+V_{2}),$ $V_{1}$ and $V_{2}$
	are defined in (\ref{eq:V1}) and (\ref{eq:V2}), respectively.
	
\end{theorem}

To estimate $V_{1}$, we can use the design-based variance estimator
applied to $m(X_{i}^{\T}\widehat{\beta})$ as
\begin{equation}
	\widehat{V}_{1}=\frac{n}{N^{2}}\sum_{i\in\mathcal{S}_{A}}\sum_{j\in\mathcal{S}_{A}}\frac{(\pi_{A,ij}-\pi_{A,i}\pi_{A,j})}{\pi_{A,ij}}\frac{m(X_{i}^{\T}\widehat{\beta})}{\pi_{A,i}}\frac{m(X_{j}^{\T}\widehat{\beta})}{\pi_{A,j}}.\label{eq:Vhat1}
\end{equation}
To estimate $V_{2},$ we further express $V_{2}$ as
\begin{equation}
	V_{2}=\frac{n}{N^{2}}\sum_{i=1}^{N}E\left[\left\{ \frac{I_{B,i}}{\pi_{B,i}(X_{i}^\T\alpha^{*})^{2}}-\frac{2I_{B,i}}{\pi_{B,i}(X_{i}^\T\alpha^{*})}\right\} \left\{ Y_{i}-m(X_{i}^{\T}\beta^{*})\right\} ^{2}+\left\{ Y_{i}-m(X_{i}^{\T}\beta^{*})\right\} ^{2}\right].\label{eq:V2-1}
\end{equation}
Let $\sigma^{2}(X_{i}^\T\beta^{*})=E\left[\left\{ Y_{i}-m(X_{i}^{\T}\beta^{*})\right\} ^{2}\right]$,
and let $\widehat{\sigma}^{2}(X_{i})$ be a consistent estimator of
$\sigma^{2}(X_{i}^\T\beta^{*})$. We can then estimate $V_{2}$ by
\[
\widehat{V}_{2}=\frac{n}{N^{2}}\sum_{i=1}^{N}\left[\left\{ \frac{I_{B,i}}{\pi_{B}(X_{i}^\T\widehat{\alpha})^{2}}-\frac{2I_{B,i}}{\pi_{B}(X_{i}^\T\widehat{\alpha})}\right\} \left\{ Y_{i}-m(X_{i}^{\T}\widehat{\beta})\right\} ^{2}+I_{A,i}d_{A,i}\widehat{\sigma}^{2}(X_{i})\right].
\]
By the law of large numbers, $\widehat{V}_{2}$ is consistent for
$V_{2}$ regardless whether one of $\pi_{B,i}(X_{i}^\T\alpha)$ or $\pi_{B,i}(X_{i}^{\T}\beta)$
is misspecificed, and therefore it is doubly robust.

\begin{theorem}[Double robustness of $\widehat{V}$]
	
	Under Assumptions \ref{asmp:MAR}\textendash \ref{assump:regularity },
	if either $\pi_{B}(X^\T\alpha)$ or $m(X^\T\beta)$ is correctly specified,
	$\widehat{V}=\widehat{V}_{1}+\widehat{V}_{2}$ is consistent for $V$.
	
\end{theorem}

\section{Simulation Study\label{sec:A-simulation-study}}

\subsection{Setup}

In this section, we evaluate the finite-sample performance of the
proposed procedure. We first generate a finite population $\mathcal{F}_{N}=\{(X_{i},Y_{i}):i=1,\ldots N\}$
with $N=10,000$, where $Y_{i}$ is a continuous or binary outcome
variable, and $X_{i}=(1,X_{1,i},\ldots,X_{p-1,i})^{\T}$ is a $p$-dimensional
vector of covariates with the first component being $1$ and other
components independently generated from standard normal with mean
$0$ and variance $1$. We set $p=50$. From the finite population,
we select a non-probability sample $\B$ of size $n_{B}\approx 2,000$,
according to the inclusion indicator $I_{B,i}\sim$Ber$(\pi_{B,i})$.
We select a probability sample $\A$ of the average size $n_{A}=500$
under Poisson sampling with $\pi_{A,i}\propto(0.25+|X_{1i}|+0.03|Y_{i}|)$.
The parameter of interest is the population mean $\mu=N^{-1}\sum_{i=1}^{N}Y_{i}$.

For the non-probability sampling probability, we consider both linear
and nonlinear sampling score models
\begin{description}
	\item [{PSM$\ $I:}] $\logit(\pi_{B,i})=\alpha_{0}^{\T}X_{i}$, where $\alpha_{0}=(-2,1,1,1,1,0,0,0,\ldots,0)^{\T}$,
	\item [{PSM$\ $II:}] $\logit(\pi_{B,i})=3.5+\alpha_{0}^{\T}\log(X_{i}^{2})-\sin(X_{3,i}+X_{4,i})-X_{5,i}-X_{6,i}$,
	where $\alpha_{0}=(0,0,0,3,3,3,3,0,\ldots,0)^{\T}$.
\end{description}


For generating a continuous outcome variable $Y_{i}$, we consider
both linear and nonlinear outcome models with $\beta_{0}=(1,0,0,1,1,1,1,0\ldots,0)^{\T}$:
\begin{description}
	\item [{OM$\ $I:}] $Y_{i}=\beta_{0}^{\T}X_{i}+\epsilon_{i},$ $\epsilon_{i}\sim\N(0,1)$,
	\item [{OM$\ $II:}] $Y_{i}=1+\exp\left\{ 3\sin\left(\beta_{0}^{\T}X_{i}\right)\right\} +X_{5,i}+X_{6,i}+\epsilon_{i},$
	$\epsilon_{i}\sim\N(0,1)$.
\end{description}

For generating a binary outcome variable $Y_{i}$, we consider both
linear and nonlinear outcome models with $\beta_{0}=(1,0,0,3,3,3,3,0\ldots,0)^{\T}$,
\begin{description}
	\item [{OM$\ $I:}] $Y\sim$Ber$\{\pi_{Y}(X)\}$ with logit$\{\pi_{Y}(X)\}=\beta_{0}^{\T}X$,
	\item [{OM$\ $II:}] $Y\sim$Ber$\{\pi_{Y}(X)\}$ with logit$\{\pi_{Y}(X)\}=2-\log\left\{ \left(\beta_{0}^{\T}X\right)^{2}\right\} +2X_{5,i}+2X_{6,i}$.
\end{description}


We consider the following estimators:
\begin{description}
	\item [{Naive,$\ \hat{\mu}_{\naive}$,}] the naive estimator using the
	simple average of $Y_{i}$ from Sample B, which provides the degree
	of the selection bias;
	\item [{Oracle,}] $\hat{\mu}_{\ora}$, the doubly robust estimator $\widehat{\mu}_{\dr}(\widehat{\alpha}_{\ora},\widehat{\beta}_{\ora}),$
	where $\widehat{\alpha}_{\ora}$ and $\widehat{\beta}_{\ora}$ are
	based on the joint estimation restricting to the known important covariates
	for comparison purpose;
	\item [{p-ipw,$\ \hat{\mu}_{\pipw}$,}] the penalized inverse probability
	of sampling weighting estimator $\hat{\mu}_{\ipw}=$ $N^{-1}\sum_{i\in\B}\widehat{\pi}_{B,i}^{-1}Y_{i},$
	where $\widehat{\pi}_{B,i}=\pr(I_{B,i}=1\mid X_{i}^{\T}\widehat{\alpha})$,
	and $\widehat{\alpha}$ is obtained by a weighted penalized regression
	of $I_{B,i}$ on $X_{i}$ using the combined sample of A and B, weighted
	by the design weights;
	\item [{p-reg,$\ \hat{\mu}_{\preg}$,}] the penalized regression estimator
	$\hat{\mu}_{\preg}=N^{-1}\sum_{i\in\A}d_{A,i}m(X;\widehat{\beta})$,
	where $\widehat{\beta}$ is obtained by a penalized regression of
	$Y_{i}$ on $X_{i}$ based on Sample B;
	\item [{p-dr0,$\ \hat{\mu}_{\pee}$,}] the penalized double estimating
	equation estimator based on the set of outcome predictors $\widehat{\mathcal{M}}_{\beta}$;
	\item [{p-dr,$\ \hat{\mu}_{\pee}$,}] the proposed penalized double estimating
	equation estimator based on the union of sampling and outcome predictors
	$\widehat{\mathcal{M}}_{\alpha}\cup\widehat{\mathcal{M}}_{\beta}$;.
\end{description}
We also note that $\widehat{\mu}_{\dr}$ without variable selection
is severely biased and unstable and therefore is excluded for comparison.

\subsection{Simulation Results}

All simulation results are based on $500$ Monte Carlo runs. Table
\ref{tab:ResultsSelection} reports the selection performance of the
proposed penalized estimating equation approach in terms of the proportion
of the proposed procedure under-selecting (Under), over-selecting
(Over), the average false negatives (FN: the average number of selected
covariates that have the true zero coefficients), and the average
false positives (FP: the average number of selected covariates that
have the true zero coefficients). The proposed procedure selects all
covariates with nonzero coefficients in both outcome model and the
sampling score model under the true model specification. Moreover,
the number of false positives is small under the true model specification.

\begin{figure}
	\begin{centering}
		\includegraphics[scale=0.7]{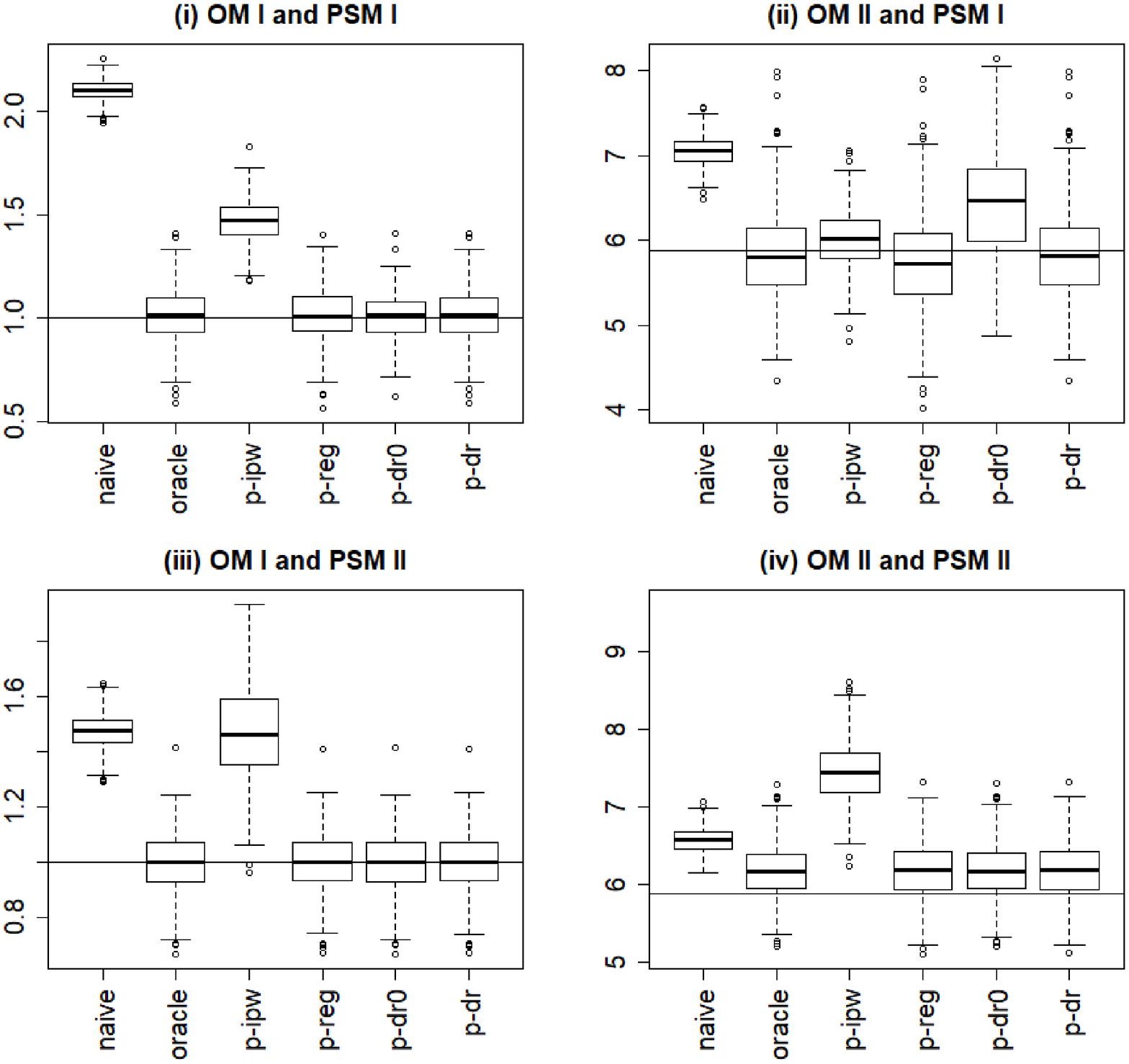}
		\par\end{centering}
	\caption{\label{fig:sim1-cross}Estimation results for the \textit{continuous
			outcome} under four scenarios: under OM I (II), the outcome model
		is correctly specified (misspecified), and under PSM I (II), the probability
		of sampling score model is correctly specified (misspecified)}
\end{figure}

Figure \ref{fig:sim1-cross} displays the estimation simulation results
for the continuous outcome. The naive estimator $\hat{\mu}_{\naive}$
shows large biases across scenarios. The oracle estimator $\hat{\mu}_{\ora}$
is doubly robust, in the sense that if either the outcome or the sampling
score is correctly specified, it is unbiased. The penalized inverse
probability of sampling weighting estimator $\hat{\mu}_{\pipw}$ shows
larges biases except for Scenario (ii). The weighted estimator $\widehat{\alpha}$
is based on the blended sample combining Sample A and Sample B, where
the units in Sample A are weighted by the known sampling weights and
the units in Sample B are weighted by $1$. This approach is justifiable
only if the sampling rate of Sample B is relatively small compared
the the population size. The penalized regression estimator $\hat{\mu}_{\preg}$
is only singly robust. When the outcome model is misspecified as in
Scenarios (ii) and (iv), it shows large biases. The proposed penalized
double estimating equation estimator $\hat{\mu}_{\pee}$ based on
$\widehat{\mathcal{M}}_{\alpha}\cup\widehat{\mathcal{M}}_{\beta}$
is doubly robust, and its performance is comparable to the oracle
estimator that requires knowing the true important variables. 
Moreover, $\hat{\mu}_{\pee}$ is slightly more efficient than $\hat{\mu}_{\ora}$.
This efficiency gain is due to using the union of covariates selected
for the sampling score model and the outcome model. This phenomenon
is consistent with the findings in \citet{brookhart2006variable}
and \citet{shortreed2017outcome}. The proposed penalized double estimating
equation estimator $\widehat{\mu}_{\mathrm{p-dr0}}$ based on $\widehat{\mathcal{M}}_{\beta}$
is slightly more efficient than $\widehat{\mu}_{\mathrm{p-dr}}$ based
on $\widehat{\mathcal{M}}_{\alpha}\cup\widehat{\mathcal{M}}_{\beta}$
in Scenario (i) when both the outcome and sampling score models are
correctly specified; however, $\widehat{\mu}_{\mathrm{p-dr0}}$ has
a large bias in Scenario (ii) when the outcome model is misspecified
and therefore is not doubly robust anymore; see Remark \ref{rmk:joint}.

\begin{figure}
	\begin{centering}
		\includegraphics[scale=0.7]{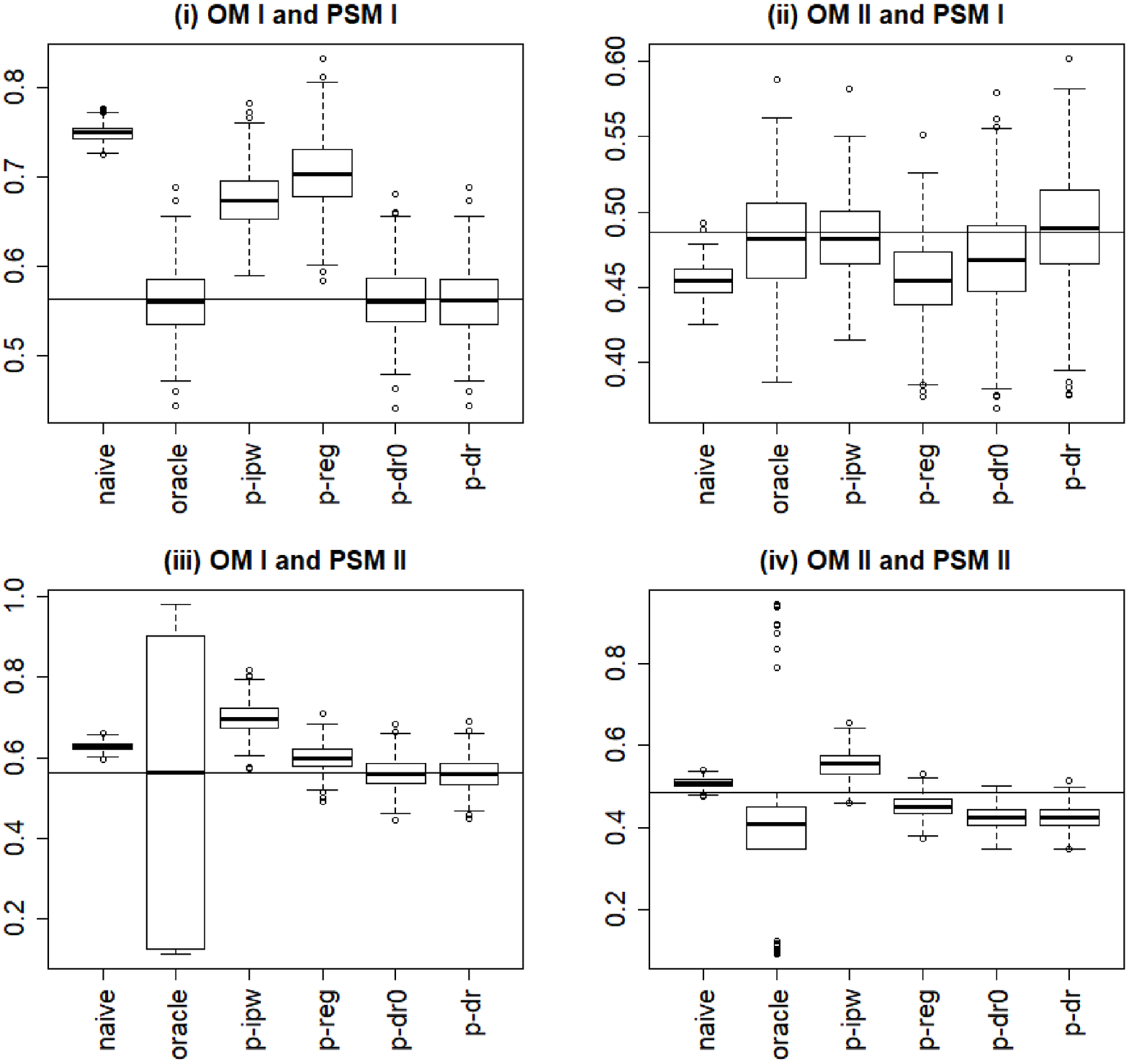}
		\par\end{centering}
	\caption{\label{fig:sim-glm-2}Estimation results for the\textit{ binary outcome}
		under four scenarios: under OM I (II), the outcome model is correctly
		specified (misspecified), and under PSM I (II), the probability of
		sampling score model is correctly specified (misspecified)}
\end{figure}

Figure \ref{fig:sim-glm-2} displays the estimation results for the
binary outcome. The same discussion above applies here. Moreover,
when the outcome model is incorrectly specified, the oracle estimator
has a large variability. In this case, the proposed estimator outperforms
the oracle estimator, because the variable selection step helps to
stabilize the estimation performance.

Table \ref{tab:ResultsCVG} reports the simulation results for the
coverage properties for the continuous outcome and binary outcome.
Under the double robustness condition (i.e., if either the outcome
model or the sampling score model is correctly specified), the coverage
rates are close to the nominal coverage; while if both models are
misspecified, the coverage rates are off the nominal coverage.

\begin{table}
	\caption{\label{tab:ResultsSelection}Simulation results for selection performance
		for the proposed double penalized estimating equation procedure under
		four scenarios: under OM I (II), the outcome model is correctly specified
		(misspecified), and under PSM I (II), the probability of sampling
		score model is correctly specified (misspecified)}
	
	\centering%
	\begin{tabular}{cccccccccc}
		\hline
		& \multicolumn{4}{c}{$\beta^{*}$} &  & \multicolumn{4}{c}{$\alpha^{*}$}\tabularnewline
		& Under  & Over  & FN  & FP  &  & Under  & Over  & FN  & FP\tabularnewline
		& $(\times10^{2})$  & $(\times10^{2})$  &  &  &  & $(\times10^{2})$  & $(\times10^{2})$  &  & \tabularnewline
		\hline
		\multicolumn{10}{c}{Continuous outcome }\tabularnewline
		(i) OM I and PSM I  & 0.0  & 31.8  & 0.0  & 1.4  &  & 0.0  & 0.0  & 0.0  & 0.0\tabularnewline
		(ii) OM II and PSM I & 70.6 & 15.0 & 0.9 & 0.2 &  & 0.0  & 0.0  & 0.0  & 0.0\tabularnewline
		(iii) OM I and PSM II  & 0.0  & 32.8  & 0.0  & 1.4  &  & 100.0  & 100.0  & 4.0  & 1.0\tabularnewline
		(iv) OM II and PSM II  & 0.0 & 0.4 & 0.0 & 0.4 &  & 100.0  & 100.0  & 3.5 & 4.3\tabularnewline
		\multicolumn{10}{c}{Binary outcome}\tabularnewline
		(i) OM I and PSM I  & 0.0  & 0.0  & 0.0  & 0.0  &  & 0.0  & 0.0  & 0.0  & 0.0\tabularnewline
		(ii) OM II and PSM I  & 100.0  & 0.0  & 2.1  & 0.0  &  & 0.0  & 0.0  & 0.0  & 0.0\tabularnewline
		(iii) OM I and PSM II  & 0.0  & 0.0  & 0.0  & 0.0  &  & 100.0  & 100.0  & 4.0  & 1.0\tabularnewline
		(iv) OM II and PSM II  & 100.0  & 0.0  & 4.0  & 0.0  &  & 100.0  & 96.0  & 4.0  & 1.0\tabularnewline
		\hline
	\end{tabular}
\end{table}

\begin{table}
	\caption{\label{tab:ResultsCVG}Simulation results for the coverage properties
		for the continuous and binary outcomes: empirical coverage rate and
		(empirical coverage rate$\pm2\times$Monte Carlo standard error)}
\begin{center}
		\begin{tabular}{cccccc}
			\hline
			& \multicolumn{2}{c}{Continuous outcome} &  & \multicolumn{2}{c}{Binary outcome}\tabularnewline
			\hline
			(i) OM I and PSM I  &  & $95.2\ (93.3,97.1)$ &  &  & $95.7\ (93.9,97.6)$\tabularnewline
			(ii) OM II and PSM I  &  & $94.6\ (92.6,96.6)$ &  &  & $95.5\ (93.6,97.4)$\tabularnewline
			(iii) OM I and PSM II  &  & $96.2\ (94.2,97.8)$ &  &  & $95.6\ (93.8,97.5)$\tabularnewline
			(iv) OM II and PSM II  &  & $88.2\ (85.3,91.1)$ &  &  & $42.9\ (38.3,47.6)$\tabularnewline
			\hline
	\end{tabular}
\end{center}
\end{table}

\section{An Application\label{sec:Real-data-application}}
\begin{figure}
	\begin{center}
		\includegraphics[scale=0.5]{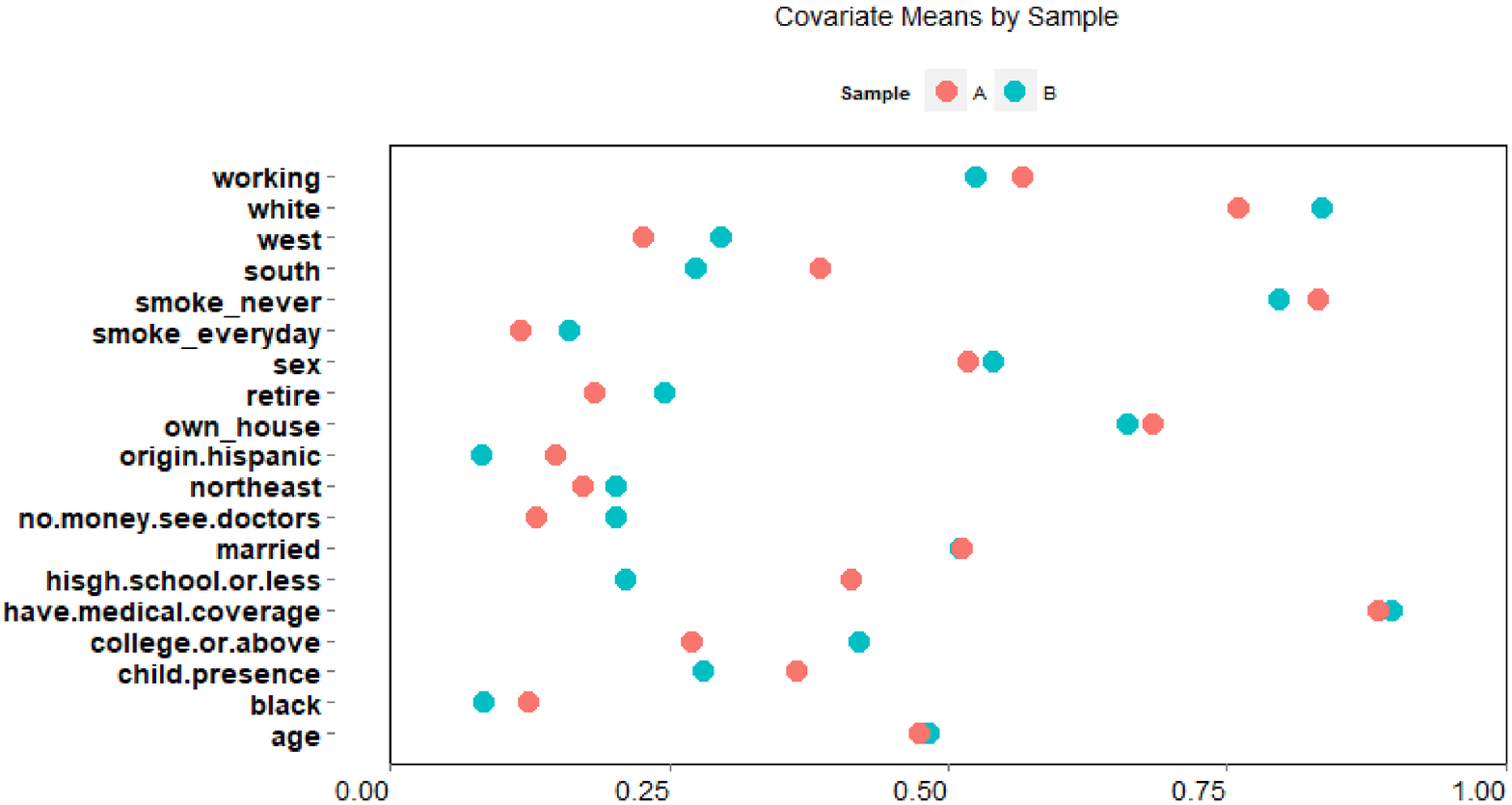}
	\end{center}
	\caption{\label{fig:real-covmeans}The covariate means by two samples: age
		is divided by $100$}
\end{figure}
We analyze two datasets from the 2005 Pew Research Centre (PRC, http://www.pewresearch.org/)
and the 2005 Behavioral Risk Factor Surveillance System (BRFSS). The
goal of the PRC study was to evaluate the relationship between individuals
and community \citep{chen2018doubly,kim2018combining}. The 2005 PRC
dataset is from a non-probability sample  provided by eight different vendors,
which consists of $n_{B}=9,301$ subjects. We focus on two study variables,
a continuous $Y_{1}$ (days had at least one drink last month) and
a binary $Y_{2}$ (an indicator of voted local elections). The 2005
BRFSS sample is a probability sample, which consists of $n_{A}=441,456$
subjects with survey weights. This dataset does not have measurements
on the study variables of interest; however, it contains a rich set
of common covariates with the PRC dataset listed in Figure \ref{fig:real-covmeans}.
To illustrate the heterogeneity in the study populations, Figure \ref{fig:real-covmeans}
contrasts the covariate means from the PRC data and the design-weighted
covariate means (i.e., the estimated population covariate means) from
the BRFSS dataset. The covariate distributions from the PRC sample
and the BRFSS sample are considerably different, e.g., age, education
(high school or less), financial status (no money to see doctors,
own house), retirement rate, and health (smoking). Therefore, 
the
naive analyses of the study variables based on the
PRC dataset are subject to selection biases.

We compute the naive and proposed estimators. To apply the proposed
method, we assume the sampling score to be a logistic regression model,
the continuous outcome to be a linear regression model, and the binary
outcome model to be a logistic regression model adjusting for all
available covariates. Using cross validation, the double selection
procedure identifies 18 important covariates (all available covariates
except for the northeast region) in the sampling score and the binary
outcome model, and it identifies 15 important covariates (all available
covariates except for black, indicator of smoking everyday, the northeast
region and the south region).

Table \ref{tab:Results-prc} presents the point estimate and the standard
error. For estimating the standard error, because the second-order
inclusion probabilities are unknown, following the survey literature,
we approximate the variance estimator in (\ref{eq:Vhat1}) by assuming the survey design is
single-stage Poisson sampling. We find significant differences in
the results between our proposed estimator and the corresponding naive
estimator. As demonstrated by the simulation in Section \ref{sec:A-simulation-study},
the naive estimator may be biased due to selection biases, and the
proposed estimator utilizes a probability sample to correct for such
biases. From the results, on average, the target population had at
least one drink for $4.84$ days over the last month, and $71.8\%$
of the target population voted in local elections.

\begin{table}[H]
	\caption{\label{tab:Results-prc}Point estimate, standard error and $95\%$
		Wald confidence interval}
	
	\centering%
	\resizebox{\textwidth}{!}{
	\begin{tabular}{cccccccc}
		\hline
		& \multicolumn{3}{c}{$Y_{1}$ (days had at least one drink last month)} &  & \multicolumn{3}{c}{$Y_{2}$ (whether voted local elections)}\tabularnewline
		& Est  & SE & CI &  & Est$\times10^{2}$ & SE$\times10^{2}$ & CI$\times10^{2}$\tabularnewline
		\hline
		Naive & 5.36  & 0.90 & (5.17,5.54) &  & 75.3 & 0.5 & (74.4,76.3)\tabularnewline
		Proposed method & 4.84 & 0.15 & (4.81,4.87) &  & 71.8  & 0.2 & (71.3,72.2)\tabularnewline
		\hline
	\end{tabular}
}
\end{table}




\section*{Acknowledgment}

Dr. Yang is partially supported by \textit{the National Science Foundation grant DMS 1811245, National Cancer Institute grant P01 CA142538, and Oak Ridge Associated Universities}. Dr. Kim is partially supported by \textit{the National Science Foundation MMS-1733572}.
An R package "IntegrativeFPM" that implements the proposed method is available at {\url{https://github.com/shuyang1987/IntegrativeFPM}}.

\bigskip
\begin{center}
	{\large\bf SUPPLEMENTARY MATERIAL}
\end{center}

Supplementary material
	 provides technical details and proofs.

\bibliographystyle{dcu}
\bibliography{DML}

\end{document}